\documentstyle[aps,prb,multicol,ifthen,epsfig]{revtex}

\newcommand{\wide}[2]{                                                                                             %
\end{multicols}                                                                                                                 %
\widetext                                                                                                                            %
\noindent                                                                                                                           %
\ifthenelse{\equal{#1}{t}}                                                                                              %
{}                                                                                                                                           %
{                                                                                                                                            %
\raisebox{0.1in}[0in][0.02in]{$\rule{3.575in}{0.002in}                                            %
\rule{0.002in}{0.08in}$}                                                                                                  %
}                                                                                                                                            %
#2                                                                                                                                         %
\ifthenelse{\equal{#1}{b}}                                                                                             %
{}                                                                                                                                           %
{                                                                                                                                            %
{\raisebox{-0.1in}[0in][0.02in]                                                                                       %
{\hspace{3.575in}$\rule{0.002in}{0.08in}                                                                   %
\rule[0.08in]{3.575in}{0.002in}$}                                                                                   %
}                                                                                                                                             %
}                                                                                                                                             %
\begin{multicols}{2}                                                                                                         %
\noindent                                                                                                                            %
}                                                                                                                                             %

\def  \bsig    {\mbox{\boldmath$\sigma$}}
\def  \brho    {\mbox{\boldmath$\rho$}}

\begin{document}
\draft

\title{Electrons in a ferromagnetic metal with a domain wall}
\author{V. K. Dugaev$^{1,}$\cite{email}, J.~Barna\'s$^{2}$,
A.~{\L}usakowski$^3$ and {\L}.~A.~Turski$^4$}
\address{$^1$Department of Electronics and Communications, Instituto Superior
de Engenharia de Lisboa,\\
Rua Conselheiro Emidio Navarro, Lisbon, Portugal, and\\
Institute for Problems of Materials Science, Ukrainian Academy of Sciences,
Vilde 5, 58001 Chernovtsy, Ukraine\\
$^2$Department of Physics, Adam Mickiewicz University,
ul. Umultowska 85, 61-614 Pozna\'n, and\\
Institute of Molecular Physics, Polish Academy of Sciences,
ul. M. Smoluchowskiego 17, 60-179 Pozna\'n, Poland\\
$^3$Institute of Physics, Polish Academy of Sciences,
Al. Lotnik\'ow 32/46, 02-668 Warsaw, Poland\\
$^4$Center of Theoretical Physics, Polish Academy of Sciences, and\\
Department of Mathematics and Natural Sciences = College of Science
Cardinal Wyszy\'nski University,\\
Al. Lotnik\'ow 32/46, 02-668 Warsaw, Poland}

\date{\today }
\maketitle

\begin{abstract}

We present theoretical description of conduction electrons interacting
with a domain wall in ferromagnetic metals.
The description takes into account interaction between electrons.
Within the semiclassical approximation we calculate the spin and charge
distributions, particularly their modification by the domain wall.
In the same approximation
we calculate local transport characteristics, including relaxation times
and charge and spin conductivities. It is shown that these parameters
are significantly modified near the wall and this modification
depends on  electron-electron interaction.

\end{abstract}
\pacs{75.60.Ch, 75.70.Cn, 75.70.Pa}

\begin{multicols}{2}

\section{Introduction}

It is generally believed that domain walls modify significantly
all the transport properties of ferromagnetic metals.
Early theoretical analysis of this phenomenon \cite{cabrera,berger}
were stimulated by magnetotransport
measurements on single-crystal Fe-wiskers.
Recently progress in controlling magnetic state of nanostructures
enabled observation of a direct correlation between domain
structure and transport properties.\cite{kent1}
For example it is
possible to extract  the contribution
to resistivity due to a single domain wall. In addition,
the discovery of Giant Magnetoresistance
(GMR) in magnetic multilayers, which is connected with reorientation of
the magnetic moments of neighboring magnetic layers from antiparallel to
parallel alignments,
renewed the interest in domain wall resistivity.
To some extent, the domain wall plays a similar role as
the nonmagnetic layer separating two ferromagnetic films in a
sandwich structure or in a multilayer,
and therefore can be expected to
lead to magnetoresistance effects similar to GMR.
Indeed, there is a growing experimental evidence
of a large magnetoresistance due to a domain wall in ferromagnetic
nanostructures.\cite{kent1,gorkom99,garcia,kent,rudiger} This, in turn,
led to growing interest in theoretical
understanding of the behavior of electrons coupled
to a ferromagnetic domain
wall.\cite{tatara97,tat99,levy,hoof,brataas,gorkom,jonkers}
Moreover, progress in nanotechnology made also possible to
study electric current flowing through
a narrow contact between two ferromagnetic metals (point contact),
where a constrained domain wall is created in the antiparallel
configuration. Such a domain wall\cite{bruno99}
has a significant influence on the transport characteristics of the
point contact.\cite{imamura}

It has been shown experimentally\cite{gregg} that the presence of a
domain wall can either increase or decrease electrical resistance
of a system.
This intriguing observation
stimulated theoretical works on understanding of the role of a domain
wall in transport properties.
Levy et al\cite{levy} developed a semiclassical model based on the mixing of
spin-majority and spin-minority transport channels by the domain wall.
This mixing results in increase of the electric resistance due to the
presence of a wall.
On the other hand,
Tatara et al\cite{tatara97} found a negative contribution, which is
due to destruction of the weak localization corrections
to conductivity by the domain wall.
Another model which may lead either to positive or negative
contribution of a wall to resistivity was developed by Gorkom
et al.\cite{gorkom}
The key point of this model is the fact,
that the wall can lead to redistribution of the
charge carriers between spin-majority and spin-minority channels.
The domain wall contribution to resistivity depends then on
the ratio of spin-majority and spin-minority relaxation times.

In this paper, we consider electrons in a ferromagnetic metal, which
interact with a domain wall. The description includes interaction between
electrons, and therefore we use a self-consistent
analysis to describe charge and spin distributions, as well as
their modification  by the domain wall. Using the Green's function technique,
we calculate the electron relaxation times in the quasiclassical
approximation. Apart from this, we also calculate
the local charge and spin conductivities. These transport parameters are
shown to be significantly modified near the domain wall, which may give rise
to new effects.

The paper is organized as follows. In Section~2 we describe the model
as well as the
transformation used to replace the system with inhomogeneous magnetization
by a system magnetized homogeneously. In Section~3 we present the
transformed Hamiltonian, generalized by including selfconsistent fields
related to electrostatic and magnetic interactions. A semiclassical
solution
of the resulting Schr\"odinger equation for electrons is also presented there.
Scattering from the wall, in the Born approximation, is calculated in Section~4.
In Section~5 we calculate, within the quasiclassical approximation, both
the spin and charge distributions in the vicinity of the
domain wall, as well as the corresponding contributions generated by
the wall. Local relaxation times are calculated in Section~6, whereas
the local charge and spin conductivities are calculated respectively
in Sections~7 and 8. Final conclusions are provided in Section~9.

\section{Model}

Consider a general  case of a ferromagnet with a
nonuniform magnetization ${\bf M}({\bf r})$.
The one-particle Hamiltonian describing conduction electrons
locally exchange-coupled
to the magnetization ${\bf M}({\bf r})$ takes the form
\begin{equation}
\label{1}
H_0=-\frac{1}{2m}\,
\psi _{\alpha }^{\dag }\, \frac{\partial ^2}
{\partial {\bf r}^2}\; \psi _{\alpha }
-J\,
\psi _{\alpha }^{\dagger }\, \bsig _{\alpha \beta }
\cdot {\bf M}({\bf r})\, \psi _{\beta }\, ,
\end{equation}
where $J$ is the exchange parameter,
$\psi _{\alpha }$ and $\psi _{\alpha}^{\dag }$
are the spinor field operators of electrons,
$\bsig =(\sigma_x, \sigma_y, \sigma_z)$ are the Pauli matrices,
and we use the units with $\hbar=1$.

The model Hamiltonian (1) will be used to describe electrons interacting
with a domain wall in a ferromagnetic metals or in
semiconductors. The domain wall will be modeled
by a magnetization profile ${\bf M}({\bf r})$.
For the sake of simplicity
we shall assume that $\vert {\bf M}({\bf r})\vert =$ const. We can
then write
\begin{equation}
\label{2}
J\, {\bf M}({\bf r})=M {\bf n}({\bf r}),
\end{equation}
where ${\bf n}({\bf r})$ is a unit vector field to be specified later,
and $M$ measured in the energy units includes the parameter $J$.

The first step of our analysis is to perform a local unitary
transformation\cite{tatara97}
\begin{equation}
\label{3}
\psi \rightarrow T({\bf r} )\, \psi,\; \; \; \; T^{\dag }({\bf r})\, T({\bf r})=1,
\end{equation}
which removes the nonhomogeneity of ${\bf M}({\bf r})$, that is
$T({\bf r} )$ transforms the second term in Eq.~(1) as
\begin{equation}
\label{4}
\psi ^{\dag }\, \bsig \cdot {\bf n}({\bf r} )\, \psi
\rightarrow \psi^{\dag }\sigma _z\psi .
\end{equation}
The transformation matrix $T({\bf r} )$ must then obey the condition
\begin{equation}
\label{5}
T^{\dag }({\bf r})\, \bsig \cdot {\bf n}({\bf r})\, T({\bf r})
=\sigma _z\, .
\end{equation}
Explicit form for such a $T({\bf r} )$ is given by\cite{falko}
\begin{equation}
\label{6}
T({\bf r})
=\frac{1}{\sqrt{2}}\left(
\sqrt{1+n_z({\bf r})}
+i\, \frac{n_y({\bf r})\, \sigma _x-n_x({\bf r})\, \sigma _y}
{\sqrt{1+n_z({\bf r})}}\right) .
\end{equation}
The transformation (3),(6) can be applied not only to a
simple domain wall, but also to other types of topological
excitations in ferromagnetic systems, for instance helicoidal
waves, skyrmions, and others.

Applying the transformation (6) to the kinetic part of the Hamiltonian (1)
one obtains
\begin{equation}
\label{7}
\psi ^{\dag }\, \frac{\partial ^2}{\partial {\bf r}^2}\, \psi \rightarrow
\psi ^{\dag }
\left(
\frac{\partial }{\partial {\bf r}}+{\bf A}({\bf r})
\right) ^2
\psi ,
\end{equation}
where the non-Abelian gauge field ${\bf A}\left( {\bf r}\right) $
is given by
\begin{equation}
\label{8}
{\bf A}({\bf r})
=T^{\dag }({\bf r})\, \frac{\partial }{\partial {\bf r}}\, T({\bf r}).
\end{equation}
According to Eq.~(6), the gauge field ${\bf A}({\bf r})$
is a matrix in the spin space.

Let us consider now a more specific case of a domain wall in a bulk
system. Assume, the wall is translationally invariant in the $x$-$y$ plane:
${\bf M}({\bf r})\rightarrow {\bf M}(z)$ and
${\bf n}({\bf r})\rightarrow {\bf n}(z)$.
For a simple domain wall with  ${\bf M}(z)$ in the plane normal to the wall,
one can parametrize the vector ${\bf n}(z)$ as
\begin{equation}
\label{9}
{\bf n}(z)=\left( \, \sin \varphi (z),\; 0,\; \cos \varphi (z)\, \right) ,
\end{equation}
where the phase $\varphi (z)$ determines the type of a domain wall.
The transformation (6) is then reduced to
\begin{equation}
\label{10}
T(z)=\frac{1}{\sqrt{2}}
\left( \sqrt{1+\cos \varphi (z)}-i\sigma _y\;
\frac{\sin \varphi (z)}{\sqrt{1+\cos \varphi (z)}} \right) ,
\end{equation}
and the gauge field assumes a simple form
\begin{equation}
\label{11}
{\bf A}(z)
=\left( \, 0,\, 0,\, -\frac{i}{2}\, \sigma _y \; \varphi ^{\prime }(z)\, \right) ,
\end{equation}
where $\varphi ^{\prime }(z)\equiv
\partial \varphi (z)/\partial z$.

\begin{figure}
\psfig{file=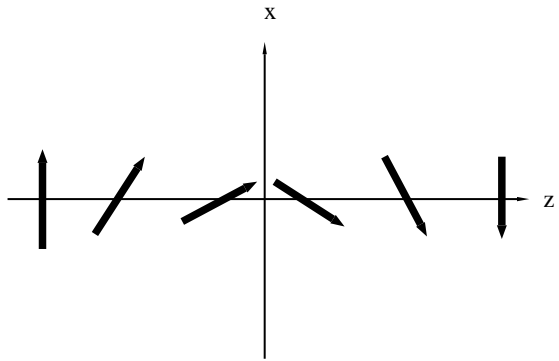,height=6cm} \caption{ Schematic picture of
the magnetization orientations near the domain wall.}
\end{figure}

Taking into account Eqs.~(7) and (11), one can write the
transformed Hamiltonian in the form:
\begin{eqnarray}
\label{12}
H_0=-\frac{1}{2m}\,
\frac{\partial ^2}{\partial {\bf r}^2}
+\frac{m\, \beta ^2(z)}{2}-M\sigma _z
\nonumber \\
+i\sigma _y\, \frac{\beta ^{\prime }(z)}{2}
+i\sigma _y \; \beta (z)\frac{\partial }{\partial z}\, ,
\end{eqnarray}
where
\begin{equation}
\label{13}
\beta (z)=\frac{\varphi ^{\prime }(z)}{2m}.
\end{equation}
For a slowly varying smooth function $\varphi (z)$ (thick domain wall
centered at $z$=0 with width $L$),
the perturbation due to the domain wall is weak, and
close to the center of the wall, $\left| z\right| \ll L$,
the parameter $\beta (z)$ can be treated as a constant.
Such a model domain wall with a constant parameter
$\beta $ was analyzed in Ref.~[\onlinecite{brataas}].

The description given above is quite general and may be used for
various models of the domain wall.
If we assume the domain wall in the form of a kink (Fig.~1), then
\begin{equation}
\label{14}
\varphi (z)=-\frac{\pi}{2}\, \tanh \, (z/L) ,
\end{equation}
and the parameter $\beta (z)$ is given by
\begin{equation}
\label{15}
\beta (z)=-\frac{\pi }{4mL\, \cosh ^2(z/L)}.
\end{equation}

\section{Semiclassical approximation}

In bulk magnetic metals like Fe, Ni or Co, the width $L$
of a magnetic domain wall is usually
much larger then the electron Fermi wavelength $\lambda _F$.
In such a case application of a semiclassical approximation is well
justified.\cite{cabrera}
The dominant perturbation from the domain wall is
then described by the term with $\beta \, (\partial /\partial z)$ in Eq.~(12),
since it is of order of $\beta k_F$.
The term proportional to $\beta^2$ is smaller, while
the term including $\beta ^{\prime }(z)$ is of the order of $\beta /L$
and therefore can be neglected.

For the sake of self-consistency, we will include now the Coulomb
interaction of electrons, which allows correct description of charge
accumulated at the wall.
The point is that the wall can give rise to
some excess charge locally breaking electrical
neutrality, as will be described in more details later.
This effect was not taken into account in previous
analysis.\cite{tatara97}
On the other hand, the renormalization of the chemical potential
forbidding the formation of excess charge\cite{brataas} may be an
over-estimation of the Coulomb repulsion.

The Coulomb interaction
will be taken into account via the coupling term
\begin{eqnarray}
\label{16}
H_{int}=\frac12 \int d^3{\bf r}
\left[ \psi ^\dag ({\bf r},t)\, \psi ({\bf r},t)-n_0\right] \,
V({\bf r}-{\bf r}^\prime) \,
\nonumber \\
\times \left[ \psi ^\dag ({\bf r}^\prime ,t)\, \psi ({\bf r}^\prime ,t)
-n_0\right] ,
\end{eqnarray}
where $V({\bf r}-{\bf r}^\prime )=e^2/\left| {\bf r}-{\bf r}^\prime \right| \, $
is the bare Coulomb interaction
and $n_0$ is the mean electron density in the bulk.
Using an auxiliary scalar field $\phi (z)$ we can incorporate the interaction
by adding to the Hamiltonian the following term (Appendix A):
\begin{equation}
\label{17} H_{int}=-\, \int d^3{\bf r}\; \phi(z)\, \psi ^{\dag }\,
\psi ,
\end{equation}
where the field $\phi(z)$ is determined by the saddle-point equation
\begin{equation}
\label{18}
\frac{d^2\phi (z)}{dz^2}=4\pi e^2\, \left( \left< \psi ^{\dag} \psi \right>
-n_0\right) ,
\end{equation}
with $\left< ..... \right>$ denoting
the ground state average.
This makes the solution self-consistent, and the field
$\phi (z)$ is the mean-field electrostatic potential in the presence
of the wall. The use of
the differential saddle-point equation for $\phi (z)$, Eq.~(18), makes the
problem more complicated due to the nonlocality,
but allows to describe correctly the screening effects associated with
a spacial distribution of charges in the vicinity of the domain wall

To include the spin-dependent interaction,
we will use a simpler formalism. More specifically,
we introduce the contact interaction term in the form
\begin{equation}
\label{19} H_{int}^s =-\frac{g_s^2}{2} \int d^3{\bf r}\left( \psi
^{\dag }\, \sigma _z\, \psi -s_0\right) ^2,
\end{equation}
where $g_s$ is the corresponding coupling constant. Choosing
$s_0$ as the spin density far from the wall guarantees that this
interaction vanishes when there is no domain wall. This means
that the effects due to magnetization of the conduction electrons
in a system without domain wall are included by the parameter $M$
in the one-particle Hamiltonian. The effect of a domain wall is
then to modify the internal magnetization, resulting from a
redistribution of the spin density. The effects due to
interaction (19) can be taken into account by adding to the
Hamiltonian a new term,
\begin{equation}
\label{20} H_{int}^{s}=\int d^3{\bf r}\; m_z(z)\, \psi ^{\dag}\,
\sigma _z\, \psi ,
\end{equation}
where the internal magnetization field $m_z(z)$ is determined
(viz. Appendix A) by
the saddle-point like equation, and is of the form
\begin{equation}
\label{21}
m_z(z)=-g_s^2\, \left( \left< \psi ^{\dag} \sigma _z\psi \right>
-s_0\right) .
\end{equation}

Thus, the Schr\"odinger equation for an electron of energy $\varepsilon $
in the fields
$\phi (z)$ and $m_z(z)$ reads:
\begin{eqnarray}
\label{22}
\left( -\frac{1}{2m}\, \frac{\partial ^2}{\partial {\bf r}^2}
+\frac{m\beta ^2(z)}{2}-[M-m_z(z)]\sigma _z
\right. \nonumber \\
\left.
+i\sigma _y \; \beta (z) \frac{\partial }{\partial z}
-\phi(z)-\varepsilon \right)\psi =0,
\end{eqnarray}
where the fields $\phi (z)$ and $m_z(z)$ have to be determined
self-consistently via Eqs. (\ref{18},\ref{21}).
Equation~(22) has the following semiclassical solutions ($i=1,2$)
\begin{eqnarray}
\label{23}
\psi _i(\brho ,z)
=\frac{\exp (\pm i\, {\bf q}\cdot \brho )}
{[\varepsilon _i^2(z)+\beta ^2(z)\, k_i^2(z)]^{1/2}\, k_i^{1/2}(z)}
\nonumber \\
\times \left(
\begin{array}{c}
\mp i\beta \, k_i(z)\\
\varepsilon _i(z)
\end{array}
\right)
\exp \left[ \pm i\int _0^zk_i(z)\, dz\right] .
\end{eqnarray}
where $\brho =(x,y)$, ${\bf q}$ is the momentum in the plane of the wall,
the wavevector components normal to the wall (along the axis $z$)
are given by
\begin{eqnarray}
\label{24}
k_{1,2}^2(z)=\kappa ^2(z)+m^2\beta ^2(z)
\nonumber \\
\pm 2m\, \left[ M_r^2(z)+\beta ^2(z)\, \kappa ^2(z)\right] ^{1/2},
\end{eqnarray}
\begin{equation}
\label{25}
\varepsilon _i(z)=\frac{k_i^2(z)}{2m}+\frac{m\beta ^2(z)}2-M_r(z)-
\frac{\kappa^2(z)}{2m},
\end{equation}
and $\kappa (z)$ and $M_r(z)$ defined as
\begin{equation}
\label{26}
\kappa ^2(z)=2m\left[ \varepsilon +\phi(z)\right] -q^2,
\end{equation}
\begin{equation}
\label{27}
M_r(z)=M-m_z(z).
\end{equation}
For clarity of notation we omitted the $z$ label of
$k_{1,2}(z)$ [$k_{z;1,2}(z)\equiv k_{1,2}(z)$].

There is no reflection from the wall in the semiclassical
approximation.
It should be noted, however, that for electrons moving nearly parallel to the
wall (with very small $z$-component of the momentum),
there is a reflection from the wall since for such electrons
the classical motion through the barrier
is impossible. We do not consider here this effect since its contribution
is very small.

\section{Scattering from the wall (Born approximation)}

For the case of not too thin domain walls,
the term proportional to $\beta (z)\, \partial /\partial z $ can
be treated as a small perturbation and
scattering from the wall can be evaluated within the Born approximation.
The matrix elements of the
$(k_z\uparrow )\rightarrow (k_z^{\prime}\downarrow)$
spin-flip scattering in the plane wave basis is given by
\begin{equation}
\label{28}
V_{k_z\, k_z^{\prime }}=\int_{-\infty }^{\infty }
{\rm e}^{-ik_z^{\prime}z}\left( -\beta (z)\,
\frac{d}{dz}\right) {\rm e}^{ik_zz}\, dz.
\end{equation}
Using (15), we obtain
\begin{equation}
\label{29}
V_{k_zk_z^{\prime }}=\frac{i\pi k_z}{4mL}
\int _{-\infty }^{\infty }
\frac{\exp \left[ -i(k_z^\prime -k_z)z\right] dz}{\cosh ^2(z/L)}.
\end{equation}
After integrating by parts, this integral can be presented as
\begin{eqnarray}
\label{29a}
V_{k_zk_z^{\prime }}=\frac{i\pi k_z}{m}
\left[ {\rm Re}\lim _{x\rightarrow \infty } {\rm e}^{-ipx}-\frac12
-p\; {\rm Im}\int _0^{\infty }\frac{{\rm e}^{-ipx}\, dx}{1+{\rm e}^{-x}}\right] ,
\nonumber
\end{eqnarray}
where $p=(k_z^\prime -k_z)L/2$. The integral in the last term can be
calculated using the series expansion of the denominator in the integrand, and
we obtain:
\begin{eqnarray}
\label{29b}
V_{k_zk_z^{\prime }}=\frac{i\pi k_z}{m}
\left[ \frac12 +p^2\sum _{n=1}^{\infty }\frac{(-1)^n}{n^2+p^2}
\right] .
\nonumber
\end{eqnarray}
Now we use the known representation\cite{abramowitz}
\begin{eqnarray}
\label{29c}
{\rm csch}(z)=\frac1{z}+2z\sum _{n=1}^{\infty }\frac{(-1)^n}{\pi ^2n^2+z^2}\; ,
\nonumber
\end{eqnarray}
and we find finally
\begin{equation}
\label{30}
V_{k_zk_z^{\prime }}
=\frac{i\pi ^2k_z(k_z^\prime -k_z)L}{4m}\;
{\rm csch}\left[ \frac{\pi (k_z^\prime -k_z)L}2\right] .
\end{equation}
Correspondingly, the probability of backscattering ($k_z^\prime =-k_z$)
is
\begin{eqnarray}
\label{30a}
W_{back}\equiv 2\pi \left| V_{k_z,-k_z}\right| ^2
=\frac{\pi ^5\, k_z^4L^2}{2m^2}\; {\rm csch}^2(\pi k_zL).
\nonumber
\end{eqnarray}
For $k_zL\gg 1$, from the last equation we find
\begin{equation}
\label{31}
W_{back}=\frac{2\pi ^5\, k_z^4L^2}{m^2}\; {\rm e}^{-2\pi k_zL}.
\end{equation}
Thus, the probability of the backscattering with simultaneous spin-flip
vanishes exponentially in the limit of $k_FL\gg 1.$\cite{cabrera}
Spin-conserving backscattering is determined by the term proportional
to $\beta^2$ in the Hamiltonian (22). In the first approximation
this term can be neglected as it
is smaller than the term proportional to $\beta (z)\, \partial /\partial z$.

The question arises, whether the Born approximation gives correct
results in the problem under consideration.
There are two general conditions for its applicability\cite{landau}
\begin{equation}
\label{32}
\left| U(z)\right| \ll \frac{1}{mL^2}\; \; \; {\rm or}\; \; \;
\left| U(z)\right| \ll \frac{k}{mL},
\end{equation}
where $U(z)$ is the scattering potential.
In the first case the Born approximation is good for arbitrary
electron energy, whereas
in the second one it is good only for fast electrons.
Therefore, if we choose the limit $k_FL\gg 1$ then
$\left| U(z)\right| \sim \beta k_z\sim (k_z/mL)$, and none of the conditions is satisfied.
In the opposite case of a small domain-wall width, $k_FL\ll 1$, we have
$\left| U(z)\right| \sim 1/mL^2$ and the Born approximation is not justified again.
Thus, the Born approximation can be used only for
rough estimations. In the case under consideration, $k_FL\gg 1$, it shows
that the usual scattering from the wall is exponentially weak.

\section{Distribution of spin and charge densities (semiclassical approach)}

In the framework of the semiclassical approximation, one can calculate
the local charge and spin densities in the vicinity of the wall, as well
as the distribution of charge and spin currents.
As follows from Section III, this can also be done taking into account
the  Coulomb interaction self-consistently.

The equation for the Green function with the term $\beta (z)$
weakly dependent on $z$,
\begin{eqnarray}
\label{33}
\left( \varepsilon +\frac{1}{2m}\, \frac{\partial ^2}{\partial {\bf r}^2}
-\frac{m\, \beta ^2(z)}{2}+\phi(z)
+M_r(z)\, \sigma _z
\right. \nonumber \\
\left.
-i\sigma _y\, \beta (z)\frac{\partial }{\partial z}+\mu \right)
G_{\varepsilon } \left( {\bf r},{\bf r}^{\prime }\right)
=\delta \left( {\bf r}-{\bf r}^{\prime }\right)\; ,
\end{eqnarray}
has a quasiclassical solution ($k_zL\ll 1$),
\begin{equation}
\label{34}
G_{\varepsilon }({\bf k})=\frac
{\varepsilon -\varepsilon _{\bf k}-M_r(z)\, \sigma _z-k_z\, \beta (z)\, \sigma _y+\mu _r }
{\left( \varepsilon -\varepsilon _{{\bf k}\uparrow }+\mu _r
+i\delta \, {\rm sgn} \, \varepsilon \right)
\left( \varepsilon -\varepsilon _{{\bf k}\downarrow }+\mu _r
+i\delta \, {\rm sgn} \, \varepsilon \right)},
\end{equation}
where the following notation has been used:
\begin{equation}
\label{35}
\varepsilon _{\bf k}=\frac{q^2+k_z^2}{2m},
\end{equation}
\begin{equation}
\label{36}
\varepsilon _{{\bf k}\uparrow ,\downarrow }
=\varepsilon _{\bf k}\mp \left[ M_r^2(z)+k_z^2\, \beta ^2(z)\right] ^{1/2},
\end{equation}
\begin{equation}
\label{37}
\mu _r=\mu -\frac{m\, \beta ^2(z)}{2}+\phi(z),
\end{equation}
and $\mu $ is the chemical potential. Equation~(36) describes the
energy spectrum in the spin-up and spin-down branches, where for the
sake of notational simplicity we droped the $z$-dependence of
$\varepsilon_{{\bf k}\uparrow ,\downarrow }$ and $\mu _r$.
In what follows we also drop the $z$-dependence of
$M_r$ and $\beta$.

Note, that the square root in Eq.~(36) contains contributions due
to spin-mixing caused by the perturbation $\sigma _y\beta \,
(\partial /\partial z)$. Hence, what we call spin-up and
spin-down branches of the spectrum (labeled as $\uparrow$ and
$\downarrow $ in Eq.~(36)) are in fact the eigenvalues
corresponding to the wavefunctions with mixed up and down states.
Correspondingly, the Green function (34) has poles at both Fermi
surfaces with $k=k_{F\uparrow }$ and $k=k_{F\downarrow }$ in
diagonal and non-diagonal components.

Using the Green function (34), one can calculate the spin density
distribution in the presence of the wall.
The real spin density distribution, i.e., transformed back to the original
basis is given by the formula
\begin{equation}
\label{38} {\bf s}(z) =
-i\; {\rm Tr}\int \frac{d\varepsilon
}{2\pi }\frac{d^3{\bf k}}{(2\pi )^3} \, T^{\dag}(z)\, \bsig \,
T(z)\, G_{\varepsilon }({\bf k}).
\end{equation}
To obtain this expression one should use the inverse of the
transformation (3), i. e. $G\rightarrow T\, G\, T^\dag $.
Using Eq.~(5), one can rewrite Eq.~(38) as
\begin{equation}
\label{39}
{\bf n}(z)\cdot {\bf s}(z)
=-i\; {\rm Tr}
\int \frac{d\varepsilon }{2\pi }\frac{d^3{\bf k}}{(2\pi )^3}\;
\sigma _z\, G_{\varepsilon }({\bf k}).
\end{equation}
Since the projection of $\, {\bf s}$ on the plane perpendicular to
$\, {\bf n}$ vanishes, we can write the spin density as
\begin{equation}
\label{40}
{\bf s}(z)
=-i\, {\bf n}(z)\, {\rm Tr}
\int \frac{d\varepsilon }{2\pi }\frac{d^3{\bf k}}{(2\pi )^3}\;
\sigma _z\, G_{\varepsilon }({\bf k}).
\end{equation}
Substituting the Green function (34) into Eq.~(40), we find
\begin{equation}
\label{41}
{\bf s}(z)={\bf n}(z) \left(
\int _{\varepsilon _{{\bf k}\uparrow }<\mu _r}
-\int _{\varepsilon _{{\bf k}\downarrow }<\mu _r}
\right)
\frac{d^3{\bf k}}{(2\pi )^3}
\frac1{\sqrt{1+k_z^2\, \beta ^2/M_r^2}}\, .
\end{equation}
After evaluating these integrals, we find the spin density distribution
\wide{m}{
\begin{eqnarray}
\label{42}
{\bf s}(z)=\frac{M_r\, {\bf n}(z)}{4\pi ^2\beta }
\left[
-\frac{k_{F\uparrow }}{2\beta }
\left( M_r^2+k_{F\uparrow }^2\beta ^2\right) ^{1/2}
+\frac{k_{F\downarrow }}{2\beta }\left(
M_r^2+k_{F\downarrow }^2\beta ^2\right) ^{1/2}
\right. \nonumber \\
\left.
+\left( 2m \mu _r+\frac{M_r^2}{2\beta ^2}\right)
\left( {\rm arcsinh}\, \frac{k_{F\uparrow }\beta }{M_r}
-{\rm arcsinh}\, \frac{k_{F\downarrow }
\beta }{M_r}\right)
+2m\beta \left( k_{F\uparrow }+k_{F\downarrow }\right) \right] ,
\end{eqnarray}
}
where
\begin{equation}
\label{43}
k_{F\uparrow ,\downarrow}^2=2m\mu _r+2m^2\beta ^2
\pm 2m\left(2m\mu _r\beta ^2+m^2\beta ^4+M_r^2\right) ^{1/2}.
\end{equation}

The accumulation of the spin density at the domain wall is
\begin{equation}
\label{44}
\Delta {\bf s}(z)={\bf s}(z)-{\bf s}_0,
\end{equation}
where ${\bf s}_0$ is the spin density in the limit of
$\beta = 0$.
For small $\beta $, i.e., for very smooth magnetic wall and up to
second order of $\beta $ this reads:
\begin{eqnarray}
\label{45}
\Delta {\bf s}(z)
=-\frac{{\bf n}(z)\, \beta ^2}{4\pi ^2}\left[
m^2\left( k_{F\uparrow }-k_{F\downarrow }\right)
\right. \nonumber \\
\left.
+\frac{m\mu }{3M_r^2}\left( k_{F\uparrow }^3-k_{F\downarrow }^3\right)
-\frac{k_{F\uparrow }^5-k_{F\downarrow }^5}{10M_r^2}\right] .
\end{eqnarray}
The sign of the factor in the square brackets
of Eq.~(45) depends on material parameters.

Charge density distribution can be calculated in a similar way,
\begin{equation}
\label{46}
\rho (z)=-i\; {\rm Tr}
\int \frac{d\varepsilon }{2\pi }\frac{d^3{\bf k}}{(2\pi )^3}\;
G_{\varepsilon }({\bf k}).
\end{equation}
After calculating the integral (46) we find
\begin{eqnarray}
\label{47}
\rho (z)=\frac{1}{4\pi ^2}\left[
2m\mu _r\left( k_{F\uparrow }+k_{F\downarrow }\right)
-\frac{k_{F\uparrow }^3+k_{F\downarrow }^3}{3}
\right. \nonumber \\
\left.
+mk_{F\uparrow }\sqrt{M_r^2+k_{F\uparrow }^2\beta ^2}
-mk_{F\downarrow }\sqrt{M_r^2+k_{F\downarrow }^2\beta ^2}
\right. \nonumber \\
\left.
+\frac{mM_r^2}{\beta }\;
{\rm arcsinh}\frac{k_{F\uparrow }\beta }{M_r}
-\frac{mM_r^2}{\beta }\;
{\rm arcsinh}\frac{k_{F\downarrow }\beta }{M_r}\right] .
\end{eqnarray}
Now we can use this expression in Eq.~(18) to determine electrostatic
potential $\phi (z)$ and $\left< \psi ^\dag \psi \right> \equiv \rho (z)$.

\begin{figure}
\psfig{file=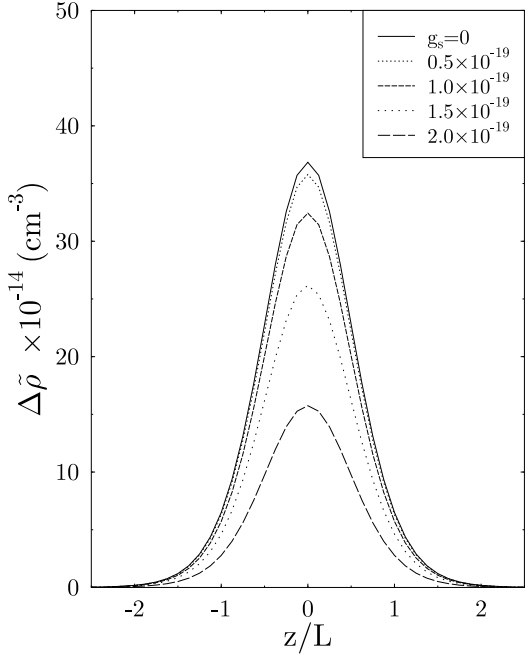,height=11cm}
\caption{Distribution of $\Delta \tilde{\rho }(z)$ near the domain wall for
different values of the spin coupling constant $g_s$. This dependence
corresponds to the charge accumulation in the absence of Coulomb repulsion.
The distribution of charge in the presence of Coulomb interactions,
$\Delta \rho (z)$, is related to $\Delta \tilde{\rho }(z)$ by Eq.~(52).
The coupling constant $g_s$ is given in units of erg$^{1/2}$cm$^{3/2}$.}
\end{figure}

We can linearize
Eq.~(18) in $\phi (z)$ assuming that the domain wall does not change
significantly the electron density, i. e. for
$\mu \gg \left| \phi (z)\right| $. Hence, after expanding $\rho (z)$ in
$\phi (z)$ and Fourier transforming over $z$, we can write Eq.~(18) as
\begin{equation}
\label{48}
\left( q_z^2+\kappa _0^2\right) \phi (q_z)
=-4\pi e^2\, \Delta \tilde{\rho }(q_z)
\end{equation}
where $\kappa _0=\left( 4\pi e^2\, \nu _0\right) ^{1/2}$ is the inverse screening
length, $\nu _0=\partial \rho /\partial \mu $ is the thermodynamic density
of states, and $\Delta \tilde{\rho }(q_z)$ is the Fourier transform of
\begin{equation}
\label{49}
\Delta \tilde{\rho }(z)\equiv \left[ \rho (z)-n_0\right] _{\, \phi (z)=0}.
\end{equation}
Using Eqs.~(27), (36), (37) and (43), we find that for $\beta \rightarrow 0$
the accumulation of charge, $\Delta \tilde{\rho}(z)=\rho- \rho(\phi =0)$  is
\begin{equation}
\label{50}
\Delta \tilde{\rho }(z)=-\frac{m^2\, \beta ^2}{4\pi ^2}
\left( k_{F\uparrow }+k_{F\downarrow }\right)
+\frac{m\beta ^2}{12\pi ^2M}\left( k_{F\uparrow }^3-k_{F\downarrow }^3\right) .
\end{equation}
This value of $\Delta \tilde{\rho }(z)$ is the accumulated charge in the absence
of Coulomb repulsion.

\begin{figure}
\psfig{file=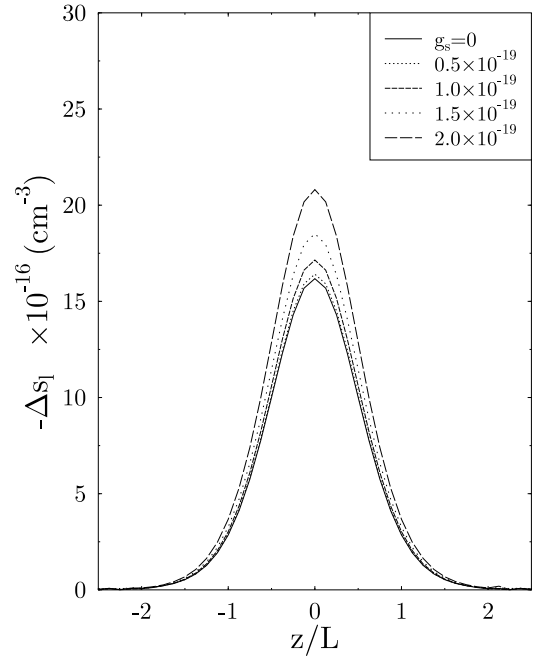,height=11cm}
\caption{Distribution of the excess spin density near the
domain wall for different values of the spin coupling constant $g_s$.}
\end{figure}

Due to (18), the real distribution of charge $\Delta \rho (z)$ is related
to $\Delta \tilde{\rho }(z)$ by the relation which in Fourier space
has the form:
\begin{equation}
\label{51}
\Delta \rho (q_z)=\frac{q_z^2}{q_z^2+\kappa _0^2}\; \Delta \tilde{\rho }(q_z).
\end{equation}

Then, if the characteristic lengh of the domain wall is large,
$\kappa _0L\gg 1$, we obtain
\begin{equation}
\label{52}
\Delta \rho (z)=-\frac1{\kappa _0^2}\, \frac{d^2\Delta \tilde{\rho }(z)}{dz^2}
\end{equation}
and, finally, using Eq.~(47), we find the distribution of charge
\begin{eqnarray}
\label{53}
\Delta \rho (z)
=-\frac{m }{2\pi ^2\kappa _0 ^2} \;
\left[ \beta (z)\frac{d^2\beta (z)}{dz^2}
+\left( \frac{d\beta (z)}{dz}\right)^2 \right]
\nonumber \\
\times \left[ -m\left( k_{F\uparrow }+k_{F\downarrow }\right)
+\frac1{3M}\left( k_{F\uparrow }^3-k_{F\downarrow }^3\right) \right].
\end{eqnarray}

We performed numerical calculations of the charge $\Delta \tilde{\rho }(z)$ and
excess spin density $\Delta s_l(z)\equiv \Delta {\bf s}(z)\cdot {\bf n}(z)$,
using the set of equations (42), (47), (49), (27), (37), (43), and (21).

\begin{figure}
\psfig{file=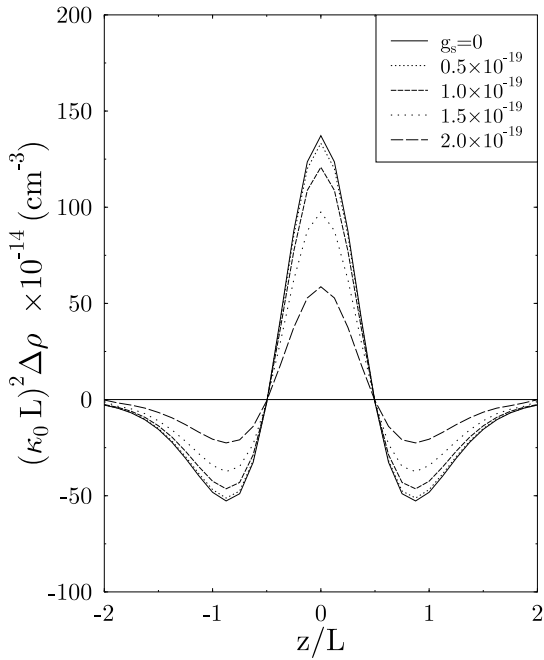,height=11cm}
\caption{Distribution of the charge density near the domain wall for
interacting electrons, calculated for different values of the spin
coupling constant $g_s$. Due to the screening effects, the integrated
charge accumulated at the wall vanishes.}
\end{figure}

The results are presented in Figs.~2 and 3 for different values of
the spin coupling constant $g_s$. In the calculations we take the
Fermi energies $\varepsilon _{F\uparrow }=3$~eV and $\varepsilon
_{F\downarrow }=2.5$~eV in the bulk, and $m=4\, m_0$, where $m_0$
is the free electron mass. The figures demonstrate how the spin
coupling constant $g_s$ affects both the spin accumulation
(Fig.~3) and $\Delta \tilde{\rho }(z)$ (Fig.~2). This effect is a
result of self-consistency, because by controlling the magnetic
density one modifies the magnetic wall, and this in turn
influences the electron density.

In view of Eq.~(49), the function $\Delta \tilde{\rho }(z)$, presented
in Fig.~2, is not the excess charge accumulated at the wall but
an auxiliary function corresponding to the condition $\phi (z)=0$.
The real distribution of accumulated charge, Eq.~(52), is presented in
Fig.~4 for different values of the coupling constant $g_s$.
This figure demonstrates that the integral of $\Delta \rho (z)$ over $z$
is zero due to the conservation of electric charge. The
characteristic length of the charge distribution is determined by the
characteristic thickness of the domain wall.

\section{Impurity self-energy}

In this Section we shall take into account the scattering of electrons
from impurities.
The simplest choice are impurities with a short-range scattering potential,
which scatter similarly both spin-up and spin-down electrons.
Let the matrix element of the scattering potential of a defect be $V_0$.
The self-energy operator in the Born approximation is
\begin{equation}
\label{54}
\Sigma (\varepsilon )=V_0^2\int \frac{d^3{\bf k}}{(2\pi )^3}\,
G_{\varepsilon }({\bf k}).
\end{equation}
After integrating the Green function given by Eq.~(34), we get
\begin{equation}
\label{55}
\Sigma (\varepsilon )=-\frac{i\; {\rm sgn}\, \varepsilon }2\; {\rm diag}\left(
1/\tau _{\uparrow },\; 1/\tau _{\downarrow }\right) ,
\end{equation}
where the momentum relaxation times for spin-up and spin-down electrons are
\begin{eqnarray}
\label{56}
\frac{1}{\tau _{\uparrow }(z)}=\frac{m\, V_0^2}{2\pi }\left(
k_{F\uparrow }+k_{F\downarrow }\hskip2cm
\right. \nonumber \\
\left.
+\frac{M_r}{\beta }\; {\rm arcsinh}\, \frac{k_{F\uparrow }\, \beta }{M_r}
-\frac{M_r}{\beta }\; {\rm arcsinh}\, \frac{k_{F\downarrow }\, \beta }{M_r}
\right) ,
\end{eqnarray}
\begin{eqnarray}
\label{57}
\frac{1}{\tau _{\downarrow }(z)}=\frac{m\, V_0^2}{2\pi }\left(
k_{F\uparrow }+k_{F\downarrow }\hskip2cm
\right. \nonumber \\
\left.
-\frac{M_r}{\beta }\; {\rm arcsinh}\, \frac{k_{F\uparrow }\, \beta }{M_r}
+\frac{M_r}{\beta }\; {\rm arcsinh}\, \frac{k_{F\downarrow }\, \beta }{M_r}\right) .
\end{eqnarray}
The difference in scattering times is due to a difference in the density
of states at the Fermi level for spin-up and spin-down electrons. The
formulae (56) and (57) take into account variation of scattering times near
the domain wall. These correlations have been neglected in previous
works.

\section{Local conductivity}

The general formula for local conductivity (without localization corrections),
when an electrical field is applied along the axis $z$, has the following form
\begin{eqnarray}
\label{58}
\sigma _{zz}=\frac{e^2}{2\pi m^2}\, {\rm Tr}\int \frac{d^3{\bf k}}{(2\pi )^3}
(k_z-m\beta \sigma _y)\; G_{\bf k}^R\;
\nonumber \\
\times (k_z-m\beta \sigma _y)\; G_{\bf k}^A,
\end{eqnarray}
where the gauge potential ${\bf A}(z)$, defined by Eq.~(11), is taken into account,
and the retarded (R) and advanced (A) Green functions are both
evaluated at the Fermi level,
\begin{eqnarray}
\label{59}
G_{\bf k}^{R,A}\equiv G_{\varepsilon =0}^{R,A}({\bf k})\hskip4cm
\nonumber \\
=\frac
{-\varepsilon _{\bf k}-M_r\sigma _z-k_z\beta \sigma _y+\mu _r}
{\left( -\varepsilon _{{\bf k}\uparrow }+\mu _r\pm i/2\tau _{\uparrow }\right)
\left( -\varepsilon _{{\bf k}\downarrow }+\mu _r\pm i/2\tau _{\downarrow }\right)}.
\end{eqnarray}

Owing to the terms containing $\beta $ in the Green functions, and
to the $z$-dependence of $\varepsilon _{k\uparrow ,\downarrow }$,
$\mu _r$ and $M_r$, the local conductivity $\sigma _{zz}$
is a smoothly varying function of $z$. Given the $\sigma _{zz}(z)$, the
resistivity of a sample with a domain wall can be found by
(adding resistivities),
\begin{equation}
\label{60}
R\sim \int \frac{dz}{\sigma _{zz}(z)}.
\end{equation}
Using Eqs.~(36) and (37), we find the local conductivity in the form
\wide{m}{
\begin{eqnarray}
\label{61}
\sigma _{zz}=\frac{e^2}{2\pi ^2m}\left[
\tau _{\uparrow }\left( \frac{k_{F\uparrow }^3}{3}
+m^2\beta ^2\, k_{F\uparrow }
-m^2M_r\beta\, {\rm arctan}\, \frac{k_{F\uparrow }\beta }{M_r}\right)
+\tau _{\downarrow }\left( \frac{k_{F\downarrow }^3}{3}
+m^2\beta ^2\, k_{F\downarrow }
-m^2M_r\beta\, {\rm arctan}\, \frac{k_{F\downarrow }\beta }{M_r}\right)
\right] .
\end{eqnarray}
One should note that the dependence on $\beta $ enters here not only
explicitly, but also through the parameters
$\tau _{\uparrow }$, $\tau _{\downarrow }$,
$k_{F\uparrow }$ and $k_{F\downarrow }$.

The description of the domain wall in terms of local conductivity is
justified when $L\gg l$, where $l$ is the electron mean free path. For
such a smooth domain wall, there is no electron scattering from the wall
but the local conductivity is changed.
The system with a domain wall is macroscopically inhomogeneous, and thus
the electric field in the vicinity of the domain wall
is inhomogeneous when a bias voltage is applied.

\section{Spin currents and local spin conductivity}

The spin-current density in the untransformed
basis has the form derived in  Appendix~ B:
\begin{equation}
\label{62}
{\bf j}_{\uparrow ,\downarrow }=-\frac{i}{2m}\, {\rm Tr}
\int \frac{d\varepsilon }{2\pi }\frac{d^3{\bf k}}{(2\pi )^3}
\left[
{\bf k}-i{\bf A}
\pm T^{\dag }\, \sigma _z\, T
\left( {\bf k}-i{\bf A}\right)
\right]
G_{\varepsilon }({\bf k}) .
\end{equation}
Suppose the spin current is induced by an electromagnetic field
with vector potential ${\bf A}_{em}$ acting on both up and
down spin components.
Then, using Eqs.~(11), (13) and (62), we obtain for the up and down spin
conductivity
\begin{eqnarray}
\label{63}
\sigma _{zz}^{\uparrow ,\downarrow }
=\frac{e}{4\pi m^2}\, {\rm Tr}\int \frac{d^3{\bf k}}{(2\pi )^3}
\left[
k_z-m\beta \sigma _y\pm (n_z\sigma _z-n_x\sigma _x)(k_z-m\beta \sigma _y\right]
\, G_{\bf k}^R \left(
k_z-m\beta \sigma _y\right) \, G_{\bf k}^A .
\end{eqnarray}
The result of calculation can be presented in the form
\begin{eqnarray}
\label{64}
\sigma _{zz}^{\uparrow ,\downarrow }
=\frac1{2e} \sigma _{zz}
+\frac{e}{4\pi ^2m}\left\{
\tau _\downarrow \left[
(m\beta ^2\mp M_rn_z)
\left(
mk_{F\downarrow }-\frac{mM_r}{\beta }\arctan \frac{k_{F\downarrow }\beta }{M_r}
\right)
\right. \right. \nonumber \\
\left. \left.
+(2m\beta ^2\mp M_rn_z)
\left(
\frac{k_{F\downarrow }}{2\beta ^2 }\sqrt{M_r^2+k_{F\downarrow }^2\beta ^2}
-\frac{M_r^2}{2\beta ^2}\, {\rm arcsinh}\, \frac{k_{F\downarrow }\beta }{M_r}
\right) \right] \right. \nonumber \\
+\tau _\uparrow \left[
(m\beta ^2\mp M_rn_z)
\left(
mk_{F\uparrow }-\frac{mM_r}{\beta }\arctan \frac{k_{F\uparrow }\beta }{M_r}
\right) \right. \nonumber \\
\left. \left.
-(2m\beta ^2\mp M_rn_z)
\left(
\frac{k_{F\uparrow }}{2\beta ^2 }\sqrt{M_r^2+k_{F\uparrow }^2\beta ^2}
-\frac{M_r^2}{2\beta ^2}\, {\rm arcsinh}\, \frac{k_{F\uparrow }\beta }{M_r}
\right) \right]
\right\} .
\end{eqnarray}
}
In the limit of $\beta \rightarrow 0$ we obtain
\begin{equation}
\label{65}
\sigma _{zz}^{\uparrow ,\downarrow }
=\frac1{2e} \sigma _{zz}
\pm \cos \varphi (z)\, \frac{e}{2m}
\left( \frac{k_{F\uparrow }\, \tau _\uparrow }{6\pi ^2}
-\frac{k_{F\downarrow }\, \tau _\downarrow }{6\pi ^2}\right) .
\end{equation}
The spin conductivity (64) and (65) describes a response in the form of up
and down spin currents to the electric field, associated with the vector
potential ${\bf A}_{em}$.

\section{Summary and conclusions}

We have described behavior of conduction electrons
interacting with a magnetic domain wall in ferromagnetic metals.
In the description we used a realistic model which includes
the Coulomb interaction and screening effects.
Within the semiclassical approximation we calculated
self-consistently the equilibrium charge and spin distribution in the
presence of a domain wall. We showed that this distribution
is significantly modified by the  wall.
We have also calculated the local transport characteristics,
like relaxation times and charge and spin conductivities.

Our approach applies to the linear response regime, and therefore such
nonequilibrium phenomena like spin accumulation at the wall
due to flowing current are not
taken into account. In a recent paper Ebels {\it et al } \cite{ebels}
observed large magnetoresistance due to a domain wall and attributed
it to the spin accumulation. On the other hand,
Simanek\cite{simanek} showed that spin accumulation is partially
suppressed by spin tracking and cannot explain such a large
magnetoresistance.

The approach used in Ref.~[\onlinecite{simanek}] is based
on the kinetic equation for the Wigner function and
takes into account nonlinear effects, particularly those due to
spin accumulation. Such effects were not included in our
description, since we analysed linear response regime only, which
is determined by equilibrium characteristics. However, we took
into account interaction between electrons and showed
that this interaction can singificanly modify
the influence of the magnetic wall on transport properties.
The local transport characteristics
were described by few parameters characterizing the domain wall.
Variation of the local conductivities at the wall may lead to
several new effects. For instance, one may expect
the Peltier effect at the domain wall.
To our knowledge, such an effect has not been studied yet.
Some other interesting phenomena may be related to
the spin dependent coupling, described by the parameter
$g_s$.

\section*{Acknowledgements}

One of the authors (V.D.) is grateful to the Mianowski Fund for the Scientific
Grant and to the Center of Theoretical Physics, Polish Academy of Sciences for the
kind hospitality.
This work is partially supported by the Polish State Committee for Scientific
Research through the Project No.~5~03B~091~20, NATO Linkage
Grant No.~977615, and NATO Science fellowship CP(UN)06/B/2001/PO.

\section*{Appendix A: Charge and spin-dependent interactions}

To include into the one-particle Hamiltonian the corrections due to
the electron-electron interactions, we use a self-consistent mean-field
approximation.
We can write down the partition function $Z$ of our system in form of
functional integral over spinor fields
$$
Z=\int D\psi ^{\dag }({\bf r},t)\; D\psi ({\bf r},t)\;
\exp \left( i\int d^3{\bf r}\, dt\, L_0\right) ,
\eqno (A.1)
$$
where $L_0$ is the Lagrangian density:
$$
\label{a2}
L_0=\psi ^{\dag} ({\bf r},t)\, \left( i\, \frac{\partial }{\partial t}-H\right)
\psi ({\bf r},t).
\eqno (A.2)
$$
The contribution to the Lagrangian from the term $H_{int}$ is
$$
\label{a3}
L_{int}=-\frac12\; \delta \rho _{\bf r}(t)\, V_{{\bf rr}^\prime }\,
\delta \rho _{{\bf r}^\prime }(t),
\eqno (A.3)
$$
where $V_{{\bf rr}^\prime }$ is an infinite matrix with the elements
$V({\bf r}-{\bf r}^\prime )$ and $\delta \rho _{\bf r}$ is a vector with
elements $\delta \rho ({\bf r})=\psi ^\dag ({\bf r},t)\, \psi ({\bf r},t)-n_0$.
Since the Coulomb interaction is instantaneous, both $\delta \rho
({\bf r},t)$ are taken at the same time $t$.

We use the Hubbard-Stratonovich method enabling decoupling of the interaction
term. It gives us the additional integration over field $\phi ({\bf r})$ in
the partition function
$$
\label{a4}
Z=\int D\psi ^{\dag }({\bf r},t)\; D\psi ({\bf r},t)\;
D\phi ({\bf r})\;
\exp \left( i\int d^3{\bf r}\, dt\, L\right) ,
\eqno (A.4)
$$
where
$$
\label{a5}
L=L_0+\phi _{\bf r}\, \delta \rho _{\bf r}
+\frac12\; \phi _{\bf r}\; V^{-1}_{{\bf rr}^\prime }\; \phi _{{\bf r}^\prime }.
\eqno (A.5)
$$
and, $\phi _{\bf r}$ is a vector constructed of the elements $\phi ({\bf r})$.
The mean field approximation corresponds to the saddle-point solution
for $\phi ({\bf r})$
$$
\label{a6}
\frac{\delta L}{\delta \phi ({\bf r})}=0,
\eqno (A.6)
$$
which gives us
$$
V^{-1}_{{\bf rr}^\prime }\, \phi _{{\bf r}^\prime }+\delta \rho _{\bf r}=0.
\eqno (A.7)
$$
After Fourier transforming and using $V({\bf q})=4\pi e^2/q^2$,
we obtain
$$
q^2\, \phi ({\bf q})+4\pi e^2\, \delta \rho ({\bf q})=0,
\eqno (A.8)
$$
and, coming back to the ${\bf r}$ - space, we get the Poisson equation
for the scalar potential
$$
\frac{\partial ^2\phi ({\bf r})}{\partial {\bf r}^2}
=4\pi e^2\, \delta \rho ({\bf r}).
\eqno (A.9)
$$

In the case of point-like interaction with
$V({\bf r}-{\bf r}^\prime ) =g\; \delta ({\bf r}-{\bf r}^\prime )$, using
the same formalism we obtain the saddle-point equation in the form
$$
\phi ({\bf r})+g\; \delta \rho ({\bf r})=0.
\eqno (A.10)
$$

\section*{Appendix B: Spin current density}

To find the expression for the spin current, we add to the Hamiltonian
of Eq.~(1) an auxiliary vector potential ${\mathcal A}_{em}(t)$
acting only on the spin-up states. It produces in
the kinetic part of the Hamiltonian the following term (in the
untransformed basis)
$$
\label{b1}
H_{kin}=
-\frac{1}{2m}\, \psi ^{\dag }({\bf r},t)
\left(
\frac{\partial }{\partial {\bf r}}
-\frac{ie}{c}\, \frac{1+\sigma _z}{2}\; {\mathcal A}_{em}(t)\right)
$$
$$
\times \left(
\frac{\partial }{\partial {\bf r}}
-\frac{ie}{c}\, \frac{1+\sigma _z}{2}\; {\mathcal A}_{em}(t)\right)
\psi ({\bf r},t).
\eqno (B.1)
$$
After expanding over ${\mathcal A}_{em}(t)$, we find the
linear in ${\mathcal A}_{em}(t)$ term in the Lagrangian density
$$
\label{b2}
\Delta L=-H_{kin}
=-\frac{ie{\mathcal A}_{em}(t)}{2mc}\;
\psi ^\dag ({\bf r},t)\; (1+\sigma _z)\;
\frac{\partial }{\partial {\bf r}}\; \psi ({\bf r},t).
\eqno (B.2)
$$
The transformation (3) changes it to
$$
\label{b3}
\Delta L=-\frac{ie{\mathcal A}_{em}(t)}{2mc}\,
\psi ^\dag ({\bf r},t)\left[
\frac{\partial }{\partial {\bf r}}
+{\bf A}({\bf r})\right.
$$
$$
\left.
+T^{\dag }({\bf r})\, \sigma _z\, T({\bf r})
\left(
\frac{\partial }{\partial {\bf r}}
+{\bf A}({\bf r})\right) \right] \psi ({\bf r},t).
\eqno (B.3)
$$
The corresponding operator of the spin-current density can be found
by variation
$$
\label{b4}
{\bf j}_{\uparrow }=\frac{c}{e}\, \frac{\delta L}
{\delta {\mathcal A}_{em}(t)},
\eqno (B.4)
$$
which gives us finally
$$
{\bf j}_\uparrow =-\frac{i}{2m}\,
\psi ^\dag ({\bf r},t)\left[
\frac{\partial }{\partial {\bf r}}
+{\bf A}({\bf r})\right.
$$
$$
\left.
+T^{\dag }({\bf r})\, \sigma _z\, T({\bf r})
\left(
\frac{\partial }{\partial {\bf r}}
+{\bf A}({\bf r})\right) \right] \psi ({\bf r},t).
\eqno (B.5)
$$

\end{multicols}

\end{document}